# Persistent Identification Of Instruments

Markus Stocker (markus.stocker@tib.eu, https://orcid.org/0000-0001-5492-3212)
TIB Leibniz Information Centre for Science and Technology, Welfengarten 1 B, 30167 Hannover, Germany and PANGAEA, Center for Marine Environmental Sciences (MARUM), University of Bremen, Leobener Str. 8, 28359 Bremen, Germany

Louise Darroch (louise.darroch@bodc.ac.uk, https://orcid.org/0000-0003-4163-9575)
British Oceanographic Data Centre, National Oceanography Centre, Liverpool, L3 5DA, United Kingdom

Rolf Krahl (rolf.krahl@helmholtz-berlin.de, https://orcid.org/0000-0002-1266-3819)
Helmholtz-Zentrum Berlin für Materialien und Energie GmbH, Albert-Einstein-Str. 15, 12489 Berlin, Germany

Ted Habermann (ted@tedhabermann.com, http://orcid.org/0000-0003-3585-6733)
Metadata Game Changers, 3980 Broadway, Suite 103-185, Boulder, Colorado, USA 80304

Anusuriya Devaraju (adevaraju@marum.de, https://orcid.org/0000-0003-0870-3192)
PANGAEA, Center for Marine Environmental Sciences (MARUM), University of Bremen, Leobener Str. 8, 28359 Bremen, Germany

Ulrich Schwardmann (ulrich.schwardmann@gwdg.de, https://orcid.org/0000-0001-6337-8674)
GWDG, Gesellschaft für wissenschaftliche Datenverarbeitung Göttingen, Am Fassberg 11, 37077 Göttingen, Germany

Claudio D'Onofrio (claudio.donofrio@nateko.lu.se, https://orcid.org/0000-0002-1982-3889)
ICOS Carbon Portal, Lund University, Physical Geography & Ecosystem Science, Sölvegatan 12, 223 62 Lund, Sweden

Ingemar Häggström (ingemar.haggstrom@eiscat.se, https://orcid.org/0000-0003-1070-6915)
EISCAT Scientific Association, Box 812, 98128 Kiruna, Sweden

# Abstract

Instruments play an essential role in creating research data. Given the importance of instruments and associated metadata to the assessment of data quality and data reuse, globally unique, persistent and resolvable identification of instruments is crucial. The Research Data Alliance Working Group Persistent Identification of Instruments (PIDINST) developed a community-driven solution for persistent identification of instruments which we present and discuss in this paper. Based on an analysis of 10 use cases, PIDINST developed a metadata schema and prototyped schema implementation with DataCite and ePIC as representative persistent identifier infrastructures and with HZB (Helmholtz-Zentrum Berlin für Materialien und Energie) and BODC (British Oceanographic Data Centre) as representative institutional instrument providers. These implementations demonstrate the viability of the proposed solution in practice. Moving forward, PIDINST will further catalyse adoption and consolidate the schema by addressing new stakeholder requirements.

# 1 Introduction

Between March 2018 and October 2019, the Research Data Alliance (RDA) Working Group (WG) Persistent Identification of Instruments (PIDINST) explored a community-driven solution for globally unambiguous and persistent identification of operational scientific instruments. By instrument, we mean *measuring instrument*, defined by the Joint Committee for Guides in Metrology (JCGM) as “device used for making measurements, alone or in conjunction with one or more supplementary devices” (VIM, 2012). Hence, PIDINST chose to address the problem of persistently identifying the devices themselves, the real-world assets with instantaneous capabilities and configurations, rather than the identification of material instrument designs (models).

Instruments are employed in numerous and diverse scientific disciplines. Instruments can be static (e.g., weather station, laboratory instrument) or mobile when mounted on moving platforms (e.g., remotely operated underwater vehicles, drones). They may be used in observation or experimentation research activities. They may be owned and operated by individual researchers, research groups, national, international or global research infrastructures or other types of institutions. For instance, at the time of writing the Integrated Carbon Observation System[1] (ICOS) operates approximately 3000 instruments at over 130 stations in 12 European countries. Astronomy is well known for their intense use of telescopes. Life Sciences employ an array of instrument types, ranging from microscopes to sequencers. Engineering Sciences, too, make heavy use of instruments.

---

[1] https://www.icos-ri.eu/

Persistent identifiers (PIDs) have a long tradition for the globally unique identification of entities relevant to or involved in research. They were developed “to address challenges arising from the distributed and disorganised nature of the internet, which often resulted in URLs to internet endpoints becoming invalid” (Klump and Huber, 2017) making it difficult to maintain a persistent record of science. Examples for well established persistent identifiers include: the Digital Object Identifier (DOI), used to identify literature, data as well as other objects (Paskin, 2009); the Open Researcher and Contributor ID (ORCID), a persistent identifier for researchers (Haak et al., 2012); the International Geo Sample Number (IGSN), a persistent identifier for physical samples and sample collections (Devaraju et al., 2016); the Research Organization Registry[2] (ROR), a persistent identifier for organizations; and the Research Resource Identifier (RRID), an identifier for physical resources, such as mice and antibodies, in the Life Sciences (Bandrowski et al., 2015).

Borgman (2015) suggested that “to interpret a digital dataset, much must be known about the hardware used to generate the data, whether sensor networks or laboratory machines”. Borgmann also highlights that “when questions arise [...] about calibration [...], they sometimes have to locate the departed student or postdoctoral fellow most closely involved”. A persistent identifier for instruments would enable research data to be persistently associated with such crucial metadata, helping to set data into context. Moreover, discovering and retrieving an instrument’s metadata through resolvable identifiers aligns with the FAIR Data Principles (Wilkinson et al., 2016), a set of guiding principles for the management of research data and its metadata. Buck et al. (2019) suggested that data provenance information is fundamental to a user’s trust in data and any data products generated. Buck et al. also recommended persistent identifiers for instruments as one of the next levels of data interoperability required to better understand and evaluate our oceans. Thus, FAIR metadata about instruments is critical in the sciences and research more broadly.

In addition to improving the FAIRness of instrument metadata, the persistent identification of instruments is also important for trusted cross-linking to valuable scientific objects, such as the research data they produce, which can be persistently identified themselves. A similar argument can be made for cross-links between instruments and literature since instruments (typically the instrument model) are generally mentioned in the literature as materials. Such cross-linking has received considerable attention in the community. The Scholix project (Burton et al., 2017) and the corresponding RDA/WDS Scholarly Link Exchange (Scholix) WG[3] have recently proposed and implemented a common schema to standardize the exchange of information about the links between literature and data. As a result, it is now easier for a data publisher that discovers a link between data and literature to share this information, and for the publisher of the article to benefit by establishing a cross-link from literature to data. With the PID Graph (Fenner and Aryani, 2019), the FREYA Project[4] is now generalizing cross-linking literature and data to other entities, including people, organizations, funders, etc. Arguably it makes good sense to enrich these connections by adding instruments.

---

[2] https://ror.org/
[3] https://www.rd-alliance.org/groups/rdawds-scholarly-link-exchange-scholix-wg
[4] https://project-freya.eu/

Currently, there is no globally implementable way to persistently identify measuring instruments. Addressing this challenge, the present article describes the results of the work conducted by PIDINST, an 18 month RDA Working Group that aimed at establishing a cost-effective, operational solution based on existing PID infrastructures, combined with a robust metadata schema for accurate identification, retrieval and automation into workflows. The solution was demonstrated at two institutional instrument providers.

# 2 Methodology

The PIDINST Case Statement[5] specified the WG objectives and deliverables. The WG took an Agile-type (empirical and iterative) approach, engaging with members and stakeholders through virtual and physical RDA Plenary meetings to ensure the results met with requirements. PIDINST operated following the methodology described in more detail in this section, summarized as follows:

- Collect use cases
- Identify common metadata
- Develop and publish the schema, and implement community feedback to its versions
- Catalyse schema implementation by existing PID infrastructure
- Prototype adoption by existing institutional instrument providers
- Engage the wider community at RDA Plenaries
- Hold regular biweekly virtual meetings.

PIDINST began with collecting use cases describing how a particular stakeholder would benefit from persistent identification of instruments. Use case descriptions included an introduction to the domain and infrastructure, (if applicable) related work by the infrastructure, and a table describing the required properties of instrument metadata associated with the persistent identifier. The metadata properties were described for their name, occurrence, definition, value datatype, and an indication whether properties should be in metadata held by the PID infrastructure or the institutional instrument provider, for instance on the landing page.

Building on the use cases, in particular the table describing the required metadata properties, PIDINST identified, organized, and harmonized the metadata properties that were common across use cases. We tabulated metadata properties as reported in use cases, harmonized their names (e.g., Identifier, Instrument Identification, and Persistent Identifier were harmonized as

5 https://rd-alliance.org/group/persistent-identification-instruments/case-statement/persistent-identification-instruments

Persistent Identifier), counted property occurrence, and grouped properties into 10 categories that emerged from the metadata analysis (i.e., were not predefined).

Given the identified common metadata, PIDINST iteratively developed a schema and obtained community feedback, particularly at RDA Plenaries. The first version was presented at the RDA 12th Plenary Meeting (Gaborone, November 2018). Following suggestions from that discussion, the properties ownerContact, ownerIdentifier, ownerIdentifierType, manufacturerIdentifier, manufacturerIdentifierType, and modelName have been added to the schema. The revised version was presented at the RDA 13th Plenary Meeting (Philadelphia, April 2019) and finally at the RDA 14th Plenary Meeting (Helsinki, October 2019). Each revision took into account community feedback at RDA Plenaries as well as issues posted on GitHub.

Having developed and published a metadata schema, PIDINST initiated discussions on schema implementation with existing PID infrastructures, in particular ePIC[6] and DataCite[7]. The discussions, held at RDA Plenaries and in virtual meetings, aimed to (1) create awareness among these infrastructures about PIDINST developments and (2) catalyse implementation. In addition to implementation by existing PID infrastructures, PIDINST also actively supported the adoption by existing institutional instrument providers through engaging institution representatives at RDA Plenaries and in virtual meetings. Several institutions have shown interest in implementing the proposed solution (Section 5) and some have already taken concrete steps (Section 3).

PIDINST had its kick-off meeting at the RDA 11th Plenary Meeting (Berlin, March 2018) and had working sessions at each subsequent Plenary until the 14th Plenary Meeting (Helsinki, October 2019) where the group had its wrap-up session. The working sessions were generally well attended with a highly engaged audience. The wider community feedback informed and validated the developments. The work was conducted between Plenaries and coordination as well as discussion was supported by biweekly open participation virtual meetings. PIDINST continues to maintain its deliverables and will be represented at future Plenaries.

# 3 Results

Between November 2017 and October 2018, the WG collected 14 use cases. An additional use case was submitted in February 2019, resulting in a total of 15 of which 14 included the table describing the required metadata and are thus considered complete. The majority of use cases are in Earth Sciences (60%). Table 1 provides an overview of the collected use cases. All use cases for which we have obtained author permission to publish are available on GitHub[8].

---

[6] https://www.pidconsortium.eu/
[7] https://datacite.org/
[8] https://github.com/rdawg-pidinst/use-cases

**Table 1:** Overview of the use cases collected by RDA WG PIDINST. *Submission* is the month of first use case submission by authors to the WG. *Completed* is the month during which the use case was completed with the required metadata. In some instances, the metadata was provided later. For UC4 the authors didn't provide the metadata; the use case was thus not completed (N/A). For UC1, UC5, UC8, and UC15 the metadata was provided after October 2018; these use cases were thus not considered in the metadata analysis.

| UC | Title | Domain | Main Author | Submission | Completed |
|---|---|---|---|---|---|
| 1 | GEOFON Global Seismic Network | Earth Sciences | Quinteros, J. | 11/2017 | 03/2019 |
| 2 | Helmholtz-Zentrum Berlin für Materialien und Energie | Multidisciplinary | Krahl, R. | 11/2017 | 06/2018 |
| 3 | National Imaging Facility, Australia | Multidisciplinary | Tapat, V. | 12/2017 | 06/2018 |
| 4 | Institute for Electromagnetic Sensing of the Environment (CNR) | Earth Sciences | Oggioni, A. | 01/2018 | N/A |
| 5 | Sensor Information System (AWI) | Earth Sciences | Macario, A. | 04/2018 | 12/2018 |
| 6 | Marine Sensor Web Enablement Working Group | Earth Sciences | Huber, R. | 05/2018 | 05/2018 |
| 7 | ORCID | Publisher | Demeranville, T. | 05/2018 | 08/2018 |
| 8 | Integrated Carbon Observation System Carbon Portal | Earth Sciences | D'Onofrio, C. | 06/2018 | 12/2018 |
| 9 | British Oceanographic Data Centre | Earth Sciences | Darroch, L. | 07/2018 | 07/2018 |
| 10 | European Southern Observatory | Astronomy | Bordelon, D. | 08/2018 | 08/2018 |
| 11 | Forschungszentrum Jülich Central Library (Journal of large-scale research facilities) | Publisher | Frick, C. | 09/2018 | 09/2018 |
| 12 | PANGAEA Data Publisher for Earth & Environmental Science | Publisher | Devaraju, A. | 09/2018 | 09/2018 |
| 13 | Permanent Service for Mean Sea Level | Earth Sciences | Darroch, L. | 10/2018 | 10/2018 |
| 14 | LTER-Europe | Earth Sciences | Oggioni, A. | 10/2018 | 10/2018 |
| 15 | UK Polar Data Centre | Earth Sciences | Tate, A. | 02/2019 | 02/2019 |

Performed in October 2018, we used the metadata of 10 then completed use cases (highlighted in Table 1) in an analysis that identified, organized, and harmonized the common properties. We tabulated properties, harmonized their names, counted property occurrence, and grouped properties into the following 10 categories: Identification, Instrument, Model, Owner, Manufacturer, Date, Capability, Output, Related Instrument, Publisher. Table 2 summarizes the analysis of metadata common to the use cases.

**Table 2:** Overview of the collected metadata, analysis of common metadata and mapping of properties onto the PIDINST schema.

| # | Property | Category | Occurrence | Schema |
|---|---|---|---|---|
| 1 | Persistent Identifier | Identification | 10 | Identifier, identifierType |
| 2 | Landing Page URL | Identification | 4 | LandingPage |
| 3 | Alternative Identifier | | 2 | AlternateIdentifier, alternateIdentifierType |
| 4 | Resource Type | | 4 | |
| 5 | Instrument Name | Instrument | 10 | Name |
| 6 | Instrument Description | Instrument | 6 | Description |
| 7 | Instrument Category | Instrument | 3 | |
| 8 | Instrument Type | Instrument | 5 | InstrumentType |
| 9 | Device URL | Instrument | 1 | |
| 10 | Model | Model | 4 | modelName |
| 11 | Sub-model | Model | 2 | |
| 12 | Instrument Owner | Owner | 6 | Owner |
| 13 | Owner Identifier | Owner | 4 | ownerIdentifier, ownerIdentifierType |
| 14 | Country | Owner | 2 | |
| 15 | Ownership Start Date | Owner | 1 | |
| 16 | Ownership End Date | Owner | 1 | |
| 17 | Contact Name | | 2 | ownerName |
| 18 | Contact eMail | | 2 | ownerContact |
| 19 | Contact Phone | | 2 | |
| 20 | Contact Institution | | 2 | |
| 21 | Institution Identifier | | 1 | |
| 22 | Manufacturer | Manufacturer | 6 | Manufacturer, manufacturerName |
| 23 | Manufacturer Identifier | Manufacturer | 1 | manufacturerIdentifier, manufacturerIdentifierType |
| 24 | Serial Number | Manufacturer | 3 | |
| 25 | Date | Date | 5 | Date |
| 26 | Date Type | Date | 2 | dateType |
| 27 | Capability | Capability | 3 | |
| 28 | Capability Type | Capability | 1 | |
| 29 | Capability Extent | Capability | 1 | |
| 30 | Characteristic | | 2 | |
| 31 | Event | | 2 | |
| 32 | Output/Observable Property | Output | 3 | VariableMeasured |
| 33 | Related Instrument Name | Related Instrument | 2 | |
| 34 | Related Instrument Identifier | Related Instrument | 1 | |
| 35 | Publisher | Publisher | 2 | |
| 36 | Publication Year | Publisher | 2 | |
| 37 | Instance Reference | | 1 | |
| 38 | Funding Reference | | 1 | |
| 39 | Related Identifier | | 1 | RelatedIdentifier |
| 40 | Related Identifier Type | | 1 | relatedIdentifierType |
| 41 | Relation Type | | 1 | relationType |
| 42 | Contributor | | 1 | |
| 43 | Contributor Type | | 1 | |

While the 43 properties collected may suggest high heterogeneity, only few can be considered common. Properties common to at least five use cases (50%) are: Persistent Identifier, Instrument Name, Instrument Description, Instrument Type, Instrument Owner, Manufacturer and Date (highlighted in Table 2). Table 2 also maps the collected properties onto the proposed PIDINST schema, which is published on GitHub[9]. As we can see, there is a mapping for all common properties. We have included additional schema properties which the WG considered important or useful even if they were not common among the considered use cases. Most notably, we include RelatedIdentifier as a flexible technique to represent identifiers of entities related to the instrument, such as articles describing the instrument or the previous version of the instrument.

PIDINST has actively supported the adoption and implementation of the schema with two stakeholders: (1) PID infrastructures, in particular DataCite and ePIC and (2) institutional instrument providers, in particular HZB (Helmholtz-Zentrum Berlin für Materialien und Energie) and BODC (British Oceanographic Data Centre).

Collaboration with DataCite resulted in a mapping of the PIDINST schema with the DataCite schema version 4.3. This mapping is also published on GitHub[10]. It shows that most instrument metadata of the PIDINST schema can be represented adequately also using the DataCite schema, even though some of the definitions need to be stretched. Still, we identified a few shortcomings with this mapping. Most notably, the DataCite schema has no suitable property for the model name. Furthermore, the controlled list of values for the resourceTypeGeneral property lacks a suitable value for Instrument. We submitted corresponding issues at the GitHub repository of the DataCite schema. Specifically, we suggest to:

- Add Instrument to the controlled list of values for resourceTypeGeneral[11]
- Add a value indicating "was used in" to relationType[12], to relate an instrument with events
- Add a Series property[13], which would solve the model name issue
- Add Name to the controlled list of values for titleType[14].

[9] https://github.com/rdawg-pidinst/schema/blob/master/schema.rst
[10] https://github.com/rdawg-pidinst/schema/blob/master/schema-datacite.rst

Amending the DataCite schema to address these issues would further increase the usability of the DataCite schema for instruments.

Collaboration with ePIC resulted in a prototypical implementation of the PIDINST schema in the ePIC infrastructure. ePIC provides a Data Type Registry infrastructure[15] that enables the definition and description of metadata schemata in a hierarchical way (Schwardmann, 2016), such that all definitions get a unique reference by a Handle. This framework is flexible enough for the definition of most possible metadata schemata. The PIDINST schema is hierarchical and contains at the first level a number of elements which contain substructures such as Owners, which is a list of objects containing ownerName, ownerContact, etc. The complete prototypical definition of the PIDINST schema is given under the name Properties-PID-instruments[16]. This definition contains all first level metadata elements of the PIDINST schema. Hence, a PIDINST metadata description can be given as a single object containing all first level metadata elements as subobjects or, for instance, as a collection of the first level metadata elements. Additionally, ePIC provides the possibility to include small metadata elements into the Handle record itself, giving useful information already at the reference level. This kind of metadata is called PID information type and is particularly useful for digital objects where metadata rather than data is of major interest.

Both the DataCite mapping and the ePIC implementation of the PIDINST schema have been prototypically tested with institutional instrument providers. As a first test case for the DataCite mapping, HZB minted four DOIs with DataCite for HZB instruments: two beamlines at the neutron source BER II[17, 18]; one beamline at the synchrotron light source BESSY II[19]; and one experimental station at BESSY II[20]. The DOIs resolve to the respective instrument page from the HZB instrument database that did already exist before and was thus not created for this purpose. One particularity with these instruments is that they are custom built by HZB. Thus, in the metadata HZB appears as Creator as well as Contributor with property contributorType value HostingInstitution. It is noteworthy that one of the DOIs uses the additional property fundingReference from the DataCite schema to acknowledge external funding that HZB received for upgrading the instrument. This property was not considered in the PIDINST schema, or in the DataCite mapping. HZB plans to continue the adoption and to mint DOIs for all its beamlines and experimental stations that are in user operation in the near future.

---

11 https://github.com/datacite/schema/issues/70
12 https://github.com/datacite/schema/issues/71
13 https://github.com/datacite/schema/issues/72
14 https://github.com/datacite/schema/issues/73
15 https://dtr.pidconsortium.net
16 https://hdl.handle.net/21.T11148/17ce618137e697852ea6
17 https://doi.org/10.5442/NI000001
18 https://doi.org/10.5442/NI000002
19 https://doi.org/10.5442/NI000003
20 https://doi.org/10.5442/NI000004

**Table 3:** Handle record of instrument identifier http://hdl.handle.net/21.T11998/0000-001A-3905-F displaying instrument metadata compliant with the PIDINST schema as implemented by ePIC.

| Type | Data |
|---|---|
| URL | https://linkedsystems.uk/system/instance/TOOL0022_2490/current/ |
| 21.T11148/8eb858ee0b12e8e463a5<br>(Identifier) | {<br>"identifierValue":"http://hdl.handle.net/21.T11998/0000-001A-3905-F",<br>"identiferType":"MeasuringInstrument"<br>} |
| 21.T11148/9a15a4735d4bda329d80<br>(LandingPage) | https://linkedsystems.uk/system/instance/TOOL0022_2490/current/ |
| 21.T11148/709a23220f2c3d64d1e1<br>(Name) | Sea-Bird SBE 37-IM MicroCAT C-T Sensor |
| 21.T11148/4eaec4bc0f1df68ab2a7<br>(Owners) | [{<br>"Owner": {<br>"ownerName":"National Oceanography Centre",<br>"ownerContact":"louise.darroch@bodc.ac.uk",<br>"ownerIdentifier":{<br>"ownerIdentifierValue":<br>"http://vocab.nerc.ac.uk/collection/B75/current/ORG00009/",<br>"ownerIdentifierType":"URL"<br>}<br>}<br>}] |
| 21.T11148/1f3e82ddf0697a497432<br>(Manufacturers) | [{<br>"Manufacturer":{<br>"manufacturerName":"Sea-Bird Scientific",<br>"modelName":"SBE 37-IM",<br>"manufacturerIdentifier":{<br>"manufacturerIdentifierValue":<br>"http://vocab.nerc.ac.uk/collection/L35/current/MAN0013/",<br>"manufacturerIdentifierType":"URL"<br>}<br>}<br>}] |
| 21.T11148/55f8ebc805e65b5b71dd<br>(Description) | A high accuracy conductivity and temperature recorder with an optional pressure sensor designed for deployment on moorings. The IM model has an inductive modem for real-time data transmission plus internal flash memory data storage. |
| 21.T11148/f76ad9d0324302fc47dd<br>(InstrumentType) | http://vocab.nerc.ac.uk/collection/L22/current/TOOL0022/ |
| 21.T11148/72928b84e060d491ee41<br>(MeasuredVariables) | [{<br>"MeasuredVariable":{<br>"VariableMeasured": |
| | "http://vocab.nerc.ac.uk/collection/P01/current/CNDCPR01/"<br>}<br>},{<br>"MeasuredVariable":{<br>"VariableMeasured":<br>"http://vocab.nerc.ac.uk/collection/P01/current/PSALPR01/"<br>}<br>},{<br>"MeasuredVariable":{<br>"VariableMeasured":<br>"http://vocab.nerc.ac.uk/collection/P01/current/TEMPPR01/"<br>}<br>},{<br>"MeasuredVariable":{<br>"VariableMeasured":<br>"http://vocab.nerc.ac.uk/collection/P01/current/PREXMCAT/"<br>}<br>}] |
| 21.T11148/22c62082a4d2d9ae2602<br>(Dates) | [{<br>"date":{<br>"date":"1999-11-01",<br>"dateType":"Commissioned"<br>}<br>}] |
| 21.T11148/eb3c713572f681e6c4c3<br>(AlternateIdentifiers) | [{<br>"AlternateIdentifier":{<br>"AlternateIdentifierValue":"2490",<br>"alternateIdentifierType":"serialNumber"<br>}<br>}] |
| 21.T11148/178fb558abc755ca7046<br>(RelatedIdentifiers) | [{<br>"RelatedIdentifier":{<br>"RelatedIdentifierValue":<br>"https://www.bodc.ac.uk/data/documents/nodb/pdf/37imbrochurejul08.pdf",<br>"RelatedIdentifierType": "URL",<br>"relationType":"IsDescribedBy "<br>}<br>}] |

BODC tested the ePIC implementation in web-published, sensor technical metadata descriptions encoded in the Open Geospatial Consortium's (OGC) SensorML[21] open standards for conceptualising and integrating real-world sensors. In an initial test case, a PID was minted for a Sea-Bird Scientific SBE37 Microcat regularly deployed on fixed-point moorings in the Porcupine Abyssal Plain Sustained Observatory (PAP-SO) in the north Atlantic[22]. As seen in the handle record[23] (Table 3), the implementation uses well-established, controlled vocabularies to facilitate adoption and (semantic) interoperability of the metadata record. The vocabularies are published by the NERC Vocabulary Server[24] (NVS), which makes use of the World Wide Web Consortium's (W3C) Simple Knowledge Organization System (SKOS) (Isaac and Summers,

21 https://www.opengeospatial.org/standards/sensorml
22 https://projects.noc.ac.uk/pap/
23 http://hdl.handle.net/21.T11998/0000-001A-3905-F?noredirect
24 https://www.bodc.ac.uk/resources/products/web_services/vocab/

2009). At BODC, each instrument's technical description is published using a unique URL[25] identifying the instrument locally, which is included in data transmissions to identify the instrument that produced them. The Handle[26] resolves directly to the instrument's technical description, which contains machine-readable metadata, such as the name, manufacturer and serial number. To enable cross-referencing, the Handle is added within the description as an identifier property labelled 'Instrument persistent identifier' (Listing 1). In this way, redirection to the instrument's SensorML URL enables globally unique identification of the instrument without costly changes to the existing publication infrastructure and data workflows.

**Listing 1:** SensorML metadata snippet showing the embedding of the instrument's (Handle) persistent identifier.

```
<sml:identifier>
  <sml:Term>
    <sml:label>Instrument persistent identifier</sml:label>
    <sml:value>http://hdl.handle.net/21.T11998/0000-001A-3904-0</sml:value>
  </sml:Term>
</sml:identifier>
```

# 4 Discussion

Rapid advances in technology means we are producing more instruments and data than ever. From simple thermistors, to large-scale synchrotrons, to global sensor observing networks, there is a growing need for innovation to address the management of these valuable assets and the data they produce. The proposed solution enables the persistent and consistent identification of instruments for citation, cross-linking and retrieval purposes across local and global instrument facilities, networks and data systems.

There are many benefits to the proposed solution. It builds on existing infrastructure and is designed to facilitate easy identification. It comprises a persistent identifier and metadata schema with a list of core metadata properties chosen for accurate identification of the instrument and for setting it into context. Metadata includes the instrument's name, a textual description, the manufacturer, the institution that owns or manages it, and references to other objects or entities that relate to the instrument. These metadata give meaning to the persistent identifier and are therefore registered with PID infrastructure.

To make persistent identification of instruments across diverse communities practical, the PIDINST schema includes only a small set of common properties. As instruments are increasingly complex and specialized, technical metadata, such as configuration and calibration, are often extensive, dynamic, and inherently difficult to standardize. There is no common standard for this kind of technical metadata that would be meaningful for all experiment techniques across scientific disciplines. However, specific standards for particular scientific communities do exist or are evolving. One approach that has had some success in Earth and Environmental observations is SensorML. Together with the Sensor Web Enablement (SWE) Common Data Model Encoding Standard, SensorML provides a conceptual model as well as XML and JSON encodings for sensors and measurement processes metadata. In general, however, the lack of standardization prevents this metadata from being registered with PID infrastructure. Instead, detailed information including descriptive material, contact information, applications, technical data, and guidance for using the instrument may be provided on the landing page associated with the instrument identifier. Technical metadata may also be linked from the metadata registered with PID infrastructure using RelatedIdentifier with property relationType value HasMetadata to enable automatic retrieval.

[25] For example, https://linkedsystems.uk/system/instance/TOOL0022_2490/current/
[26] http://hdl.handle.net/21.T11998/0000-001A-3905-F

The PIDINST schema is designed to complement multidisciplinary best practices for property values. Many properties allow for soft-typing, giving users the ability to use values of their choice, such as free text or domain-specific standards. Property attributes enable users and machines to understand the context of the value (e.g., ownerIdentifier, ownerIdentifierType), again using free text or standards. A similar approach is used in the DataCite metadata schema. Domain-specific standards can vary among communities. For example, the SeaDataNet research infrastructure and SWE Marine Profiles group recommend controlled vocabularies and identifiers to annotate datasets and open standards related to instruments (Kokkinaki et al., 2019), including the SeaVox Device Catalogue for instrument model designs, the BODC Parameter Usage Vocabulary for measured variables and the European Directory of Marine Organisations (EDMO). Communities in Earth Sciences have chosen to label measured parameters with Climate Forecast Standard Names. The PIDINST schema allows these communities to use property values of their choice. While soft-typing is practical towards multidisciplinary use it does reduce interoperability because different communities use different standards for values as described above. However, with such heterogeneity and establishment, it is impractical to use one standard for all use cases. The use of identifiers with knowledge representation schemes (e.g., SKOS) goes some way to improving understanding between information systems and can be used in the schema as shown in Listing 1. Thus, the PIDINST schema complements multidisciplinary best practices through soft-typing while facilitating the use of standards which can enhance interoperability if desired.

Another important goal for the PIDINST schema is to facilitate linking among instruments and journal articles, datasets, and other research objects. These links are made using RelatedIdentifier elements in the identifier metadata and the relationships are described using relationType elements. For instance, the identifier metadata of four instruments registered with DataCite[15,16,17,18] contain related identifiers with property relationType value IsDescribedBy. Listing 2 provides an example. The relations point to journal articles that describe the instruments and provide technical details. These articles serve a similar purpose as “data papers” (Candela et al., 2015), i.e. articles that describe datasets, published in peer reviewed journals to provide recognition for dataset creation by means of an article. Thus, we term such articles “instrument papers”. To name an example, the Journal of large-scale research facilities[27] (JLSFR) publishes such articles.

**Listing 2:** Snippet of DataCite metadata (https://api.datacite.org/dois/10.5442/NI000001) relating the HZB instrument “E2 - Flat-Cone Diffractometer” (https://doi.org/10.5442/NI000001) with a journal article describing the instrument (https://doi.org/10.17815/jlsrf-4-110).

```
"relatedIdentifiers": [{
  "relationType":"IsDescribedBy",
  "relatedIdentifier":"10.17815/jlsrf-4-110",
  “relatedIdentifierType":"DOI"
}]
```

Together, instrument papers and landing pages provide important documentation that helps scientists and users more generally understand the instruments and how they have been used in scientific experiments. This documentation is designed to be read by humans. Structured metadata linked using persistent identifiers, on the other hand, enable machine readability and processing of information about instruments. These representations of information, for humans and for machines, are complementary.

As we presented here, PIDINST prototyped the schema implementation with both DataCite and ePIC. These implementations have pros and cons which we briefly discuss. Worldwide, DOIs arguably have better recognition. Furthermore, the infrastructure for minting DataCite DOIs is easier at hand for many institutions and comes with substantial tooling. On the other hand, DataCite DOIs may incur considerable costs if DOIs are minted for a large number of instruments. Furthermore, ePIC handles are more flexible when it comes to supporting custom metadata standards. Indeed, as our results clearly demonstrate, we could implement the PIDINST schema as proposed only with ePIC.

While the PIDINST schema has already been shown to be viable in practice, it is not yet finalized in all details. One of the remaining open issues is that the group did not achieve a consensus on the best representation of an instrument's serial number. We do have the AlternateIdentifier property, so in principle, adding an AlternateIdentifier with alternateIdentifierType value SerialNumber would be the obvious way to include the serial number in the metadata. Consequently, adding a dedicated SerialNumber property to the schema has been rejected as redundant with AlternateIdentifier. The only drawback is that alternateIdentifierType is defined as free text and not a controlled vocabulary. As a result, there is no guarantee that everyone who registers instrument metadata spells this type alike, which may be a problem when searching for instrument metadata by serial number. Changing the definition of alternateIdentifierType to a controlled vocabulary is problematic, too, as there may be new use cases for AlternateIdentifier which would not be readily supported. Finally, the definition for some schema properties (e.g., InstrumentType and MeasuredVariable) is rather vague and the value is defined as free text. This is mainly due to the lack of suitable vocabularies. Other open issues with the schema, such as the final definitions of the terms in some controlled vocabularies, are relatively minor.

[27] https://jlsrf.org

A further limitation of the presented work is the relatively small set of use cases and their bias for Earth Sciences (60%), and therefore limited coverage of the disciplines. In current and future work, PIDINST will continue maintaining its deliverables, including advancing and supporting further adoptions in disciplines other than Earth Sciences. As such, PIDINST will test the viability of the proposed metadata schema and its implementation with PID infrastructures more widely. As part of metadata schema maintenance, PIDINST will consider, discuss and implement concerns different communities may have.

The use of persistent identifiers for instruments is currently an emerging solution that is gaining momentum operationally as evidenced in our adoption cases below. Increasing it's uptake in the future may also involve engagement with instrument manufacturers. They could provide machine readable instrument specifications, support including the persistent identifier into instrument output, or even register instruments "at birth".

We have recently seen the development of dedicated community-level sensor registries aimed at harmonizing and standardizing sensor metadata across instrument networks, for example, the European Esonet Yellow Pages[28] for deep sea observatories or the NSF/EarthCube X-Domes (Fredericks and Botts, 2018) for cross-domain environmental sensors. The use of persistent identifiers in such registries would not only boost uptake but these facilities may also become direct members of PID providers, minting identifiers for institutions or individuals who do not have dedicated PID services. Uptake may also be accelerated through adoption by other PID providers (e.g., EZID[29]). To support adoption in communities, PIDINST has published the schema on GitHub as a 'living document' where users may request updates to the schema, helping it to evolve with new and specialised stakeholders.

# 5 Adoption

In addition to HZB and BODC who have already demonstrated the practical viability of the proposed solution, in this section we briefly present how other research infrastructures and institutional instrument providers motivate and plan the implementation of the proposed solution.

The Integrated Carbon Observation System (ICOS) is a pan-european research infrastructure for quantifying and understanding the greenhouse gas balance of the European continent. It conducts many continuous in-situ measurements like gas concentrations, wind speed and direction, humidity, temperature, etc. To deliver high quality measurement data, ICOS considers the adoption of a persistent identifier for instruments a must for documenting data provenance and tracking calibration history.

[28] https://www.esonetyellowpages.com/
[29] https://ezid.cdlib.org/

PANGAEA[30] is a data infrastructure for archiving and publishing Earth and Environmental datasets. It is jointly managed by the Alfred Wegener Institute, Helmholtz Center for Polar and Marine Research (AWI) and the Center for Marine Environmental Sciences (MARUM) of University of Bremen. The infrastructure holds more than 380000 persistently identified (DOI) datasets from individual researchers, projects, data centers and research infrastructures. The metadata of a dataset includes relations between the dataset and related persistently identified entities such as specimens, authors, articles. Metadata can be further enriched with instrument information. Using the AWI Sensor Information System[31], a subset of the published datasets has already been linked to their instruments. Since source information of a dataset (e.g., instrument and method) is essential to interpret the quality of the dataset and to facilitate its reusability, further work should be done to link the remaining and new data submissions with their instrument PIDs, where applicable. As a data provider, PANGAEA only curates limited information of a device[32], such as device name, identifier and type. For both the persistent identification as well as for the description of instruments, PANGAEA thus relies on institutional instrument providers.

EISCAT3D[33] will be an international research infrastructure, using radar observations and the incoherent scatter technique for studies of the atmosphere and near-Earth space environment above the Fenno-Scandinavian Arctic as well as for the support of the solar system and radio astronomy sciences. EISCAT3D will implement persistent identification for instruments following the recommendations by PIDINST. The radar is complex, more digital than previous radars, and is roughly divided into a number of separate units. While software is a substantial constituent of these units, they can be regarded as hardware units, each persistently identified. Updates to the units will be primarily to software and result in new unit versions with own PIDs. The radar itself can also be persistently identified and the relation type HasComponent can be used to relate to the persistently identified units.

# 6 Conclusion

The Research Data Alliance Working Group Persistent Identification of Instruments (RDA WG PIDINST) was created with the aim to develop a community-driven solution for persistent identification of instruments. Based on use cases, the WG published a metadata schema and prototyped schema implementation with ePIC and DataCite as well as with two institutional instrument providers. The WG has thus demonstrated the practical viability of the proposed solution for persistent and consistent identification of instruments for citation, cross-linking and retrieval purposes across local and global instrument facilities, networks and data systems. We argue that one of the key advantages of the proposed solution is that it builds on existing PID infrastructure. PIDINST encourages communities to explore both the DataCite and ePIC implementation in order to gain a better understanding for which use cases they serve best. In addition to maintaining the schema and addressing new stakeholder requirements, PIDINST will continue to actively engage with stakeholders to promote further adoptions.

[30] https://pangaea.de/
[31] https://sensor.awi.de/
[32] https://github.com/rdawg-pidinst/use-cases/blob/master/12-PANGAEA.md
[33] https://eiscat.se/business/eiscat3d7/

**Acknowledgements.** We would like to thank all use case authors for their invaluable contribution to this work as well as the participants of online and offline meetings for their contributions. We also thank DataCite and ePIC for helpful cooperation. Finally, we thank the Research Data Alliance for endorsing and supporting the PIDINST WG. This work was supported by the FREYA project which has received funding from the European Union's Horizon 2020 research and innovation programme under grant agreement Nr. 777523.

# References

Bandrowski, A., Brush, M., Grethe, J. S., Haendel, M. A., Kennedy, D. N., … Hill, S. (2015). The Resource Identification Initiative: A cultural shift in publishing. F1000Research, 4, 134. https://doi.org/10.12688/f1000research.6555.2

Borgman, C.L., 2015. Big Data, Little Data, No Data. MIT Press.

Buck, J. J. H., Bainbridge, S. J., Burger, E. F., Kraberg, A. C., Casari, M., Casey, K. S., … Schewe, I. (2019). Ocean Data Product Integration Through Innovation-The Next Level of Data Interoperability. Frontiers in Marine Science, 6. https://doi.org/10.3389/fmars.2019.00032

Burton, A., Aryani, A., Koers, H., Manghi, P., La Bruzzo, S., Stocker, M., … Fenner, M. (2017). The Scholix Framework for Interoperability in Data-Literature Information Exchange. D-Lib Magazine, 23(1/2). https://doi.org/10.1045/january2017-burton (http://scholix.org)

Candela, L., Castelli, D., Manghi, P., Tani, A. (2015). Data journals: A survey. Journal of the Association for Information Science and Technology, 66(9), 1747–1762. https://doi.org/10.1002/asi.23358

Devaraju, A., Klump, J., Cox, S. J. D., Golodoniuc, P. (2016). Representing and publishing physical sample descriptions. Computers & Geosciences, 96, 1–10. https://doi.org/10.1016/j.cageo.2016.07.018

Fenner, M., Aryani, A. (2019). Introducing the PID Graph (Version 1.0). DataCite Blog. https://doi.org/10.5438/JWVF-8A66

Fredericks, J., Botts, M. (2018). Promoting the capture of sensor data provenance: a role-based approach to enable data quality assessment, sensor management and interoperability. Open Geospatial Data, Software and Standards, 3(1). https://doi.org/10.1186/s40965-018-0048-5

Haak, L.L. et al., 2012. ORCID: a system to uniquely identify researchers. Learned Publishing, 25(4), pp.259–264. https://doi.org/10.1087/20120404

International Vocabulary of Metrology - Basic and General Concepts and Associated Terms (VIM 3rd edition, 2012). https://www.bipm.org/en/publications/guides/#vim

Isaac, A.., Summers, E. (eds), SKOS Simple Knowledge Organization System Primer. W3C Working Group Note 18 August 2009. https://www.w3.org/TR/skos-primer/

Klump, J., Huber, R., 2017. 20 Years of Persistent Identifiers – Which Systems are Here to Stay?. Data Science Journal, 16, p.9. https://doi.org/10.5334/dsj-2017-009

Kokkinaki, A., Darroch, L., Buck, J., Jirka, S., 2016. Semantically Enhancing SensorML with Controlled Vocabularies in the Marine Domain. Proceedings of the Geospatial Sensor Webs Conference, Munster, Germany, August 29-31. CEUR Workshop Proceedings, vol-1762. http://ceur-ws.org/Vol-1762/Kokkinaki.pdf

Paskin, N. (2009). Digital Object Identifier (DOI®) System. Encyclopedia of Library and Information Sciences, Third Edition, 1586–1592. https://doi.org/10.1081/e-elis3-120044418

Schwardmann, U. (2016). Automated schema extraction for PID information types. 2016 IEEE International Conference on Big Data (Big Data). https://doi.org/10.1109/bigdata.2016.7840957 (Pre-print: https://hdl.handle.net/21.11101/0000-0002-A987-7)

Wilkinson, M. D., Dumontier, M., Aalbersberg, Ij. J., Appleton, G., Axton, M., Baak, A., … Bourne, P. E. (2016). The FAIR Guiding Principles for scientific data management and stewardship. Scientific Data, 3(1). https://doi.org/10.1038/sdata.2016.18